\documentclass[sigconf]{acmart}

\usepackage{hyperref}  
\usepackage{hyperxmp}
\usepackage{booktabs}
\usepackage{multirow}
\usepackage{array}
\usepackage{rotating}
\usepackage{ulem}
\usepackage{subcaption} 
\usepackage{longtable}
\usepackage{enumitem}
\usepackage{soul}
\usepackage{makecell} 

\AtBeginDocument{%
  }

\setcopyright{acmcopyright}
\copyrightyear{2025}
\acmYear{2025}
\acmDOI{XXXXXXX.XXXXXXX}

\acmConference[CHI '25]{Proceedings of the CHI Conference on Human Factors in Computing Systems}{April 26--May 1, 2025}{Yokohama, Japan}
\acmBooktitle{CHI Conference on Human Factors in Computing Systems (CHI '25), April 26--May 1, 2025, Yokohama, Japan} 

\acmSubmissionID{3320}

\begin{document}

\newcommand{\rp}[1]{\textcolor{green}{#1}}
\newcommand{\enote}[1]{\textcolor{red}{#1}}

\definecolor{lightgray}{RGB}{211, 211, 211}
\definecolor{lightblue}{RGB}{135, 206, 235}
\definecolor{lightgreen}{RGB}{144, 238, 144}
\definecolor{lightyellow}{RGB}{255, 166, 0}
\definecolor{lightpink}{RGB}{255, 182, 193}

\newcommand{\cttext}[1]{\colorbox{lightgray}{#1}}
\newcommand{\adtext}[1]{\colorbox{lightpink}{#1}}
\newcommand{\roletext}[1]{\colorbox{lightblue}{#1}}
\newcommand{\limittext}[1]{\colorbox{lightyellow}{#1}}

\newcommand{\ct}[1]{\colorunderline{lightgray}{#1}}
\newcommand{\colorunderline}[2]{\textcolor{#1}{\uline{\textcolor{black}{#2}}}}
\newcommand{\ad}[1]{\textbf{\colorunderline{lightpink}{#1}}}
\newcommand{\role}[1]{\textbf{\colorunderline{lightblue}{#1}}}
\newcommand{\limit}[1]{\textbf{\colorunderline{lightyellow}{#1}}}

\definecolor{labbg}{HTML}{27AE60}

\newcommand{\generalLabel}[1]{
    \begin{picture}(7,8) 
      \put(3,3.5){\color{labbg}\circle*{11}}
      \put(3.2,4){\makebox(0,0){{\textcolor{white}{\scriptsize\bfseries\rmfamily #1}}}} 
    \end{picture}
}

\definecolor{ppurple}{HTML}{9B59B6}

\newcommand{\sublabel}[1]{
    \begin{picture}(7,8) 
      \put(3,3.5){\color{ppurple}\circle*{11}}
      \put(3.2,4){\makebox(0,0){{\textcolor{white}{\scriptsize\bfseries\rmfamily #1}}}} 
    \end{picture}
}

\newcommand{\rr}[1]{\textcolor{black}{#1}}


\title[Understanding the LLM-ification of CHI]{Understanding the LLM-ification of CHI: Unpacking the Impact of LLMs at CHI through a Systematic Literature Review}

\author{Rock Yuren Pang}
\email{ypang2@cs.washington.edu}
\affiliation{%
 \institution{University of Washington}
 \city{Seattle}
 \state{Washington}
 \country{USA}
}

\author{Hope Schroeder}
\email{hopes@mit.edu}
\affiliation{%
 \institution{MIT}
 \city{Cambridge}
 \state{Massachusetts}
 \country{USA}
}

\author{Kynnedy Simone Smith}
\email{kynnedysimonesmith@gmail.com}
\affiliation{%
 \institution{Columbia University}
 \city{New York}
 \state{New York}
 \country{USA}
}

\author{Solon Barocas}
\email{solon@microsoft.com}
\affiliation{%
 \institution{Microsoft Research}
 \city{New York}
 \state{New York}
 \country{USA}
}

\author{Ziang Xiao}
\email{ziang.xiao@jhu.edu}
\affiliation{%
 \institution{Johns Hopkins University}
 \city{Baltimore}
 \state{Maryland}
 \country{USA}
}

\author{Emily Tseng}
\email{etseng42@gmail.com}
\affiliation{%
 \institution{Microsoft Research}
 \city{New York}
 \state{New York}
 \country{USA}
}

\author{Danielle Bragg}
\email{danielle.bragg@microsoft.com}
\affiliation{%
 \institution{Microsoft Research}
 \city{New York}
 \state{New York}
 \country{USA}
}

\renewcommand{\shortauthors}{Pang et al.}

\begin{abstract}
 Large language models (LLMs) have been positioned to revolutionize HCI, by reshaping not only the interfaces, design patterns, and sociotechnical systems that we study, but also the research practices we use. To-date, however, there has been little understanding of LLMs' uptake in HCI. We address this gap via a systematic literature review of 153 CHI papers from 2020-24 that engage with LLMs. We taxonomize: (1) domains where LLMs are applied; (2) roles of LLMs in HCI projects; (3) contribution types; and (4) acknowledged limitations and risks. We find LLM work in 10 diverse domains, primarily via empirical and artifact contributions. Authors use LLMs in five distinct roles, including as research tools or simulated users. Still, authors often raise validity and reproducibility concerns, and overwhelmingly study closed models. We outline opportunities to improve HCI research with and on LLMs, and provide guiding questions for researchers to consider the validity and appropriateness of LLM-related work.
\end{abstract}




\maketitle

\section{Introduction}

Large language models (LLMs) are poised to transform the landscape of Human-Computer Interaction (HCI) research. Already, researchers have been using LLMs across the HCI research pipeline, from ideation and system development to data analysis and paper-writing~\cite{kapania2024imcategorizingllmproductivity}. Past work has shown rapid growth in the raw count of LLM-focused paper preprints, especially in HCI topics~\cite{movva2024topics}. 
The explosion of LLM-related research has also led to rising discourse in HCI on the opportunities and challenges of LLM usage, including interview and survey studies with researchers to understand their practices \cite{kapania2024imcategorizingllmproductivity}, and workshops \cite{aubin2024llm, prpa2024challenges} and social media commentary~\cite{Hullman2024} in which scholars debate how the field ought to respond.
The surge in LLM-related papers and discussions indicates a growing need to support scholars in understanding the potential and pitfalls of LLMs in HCI.

Such inquiry is consequential not only for HCI, but also for the broader landscape of computing research.
On one hand, scholars in natural language processing (NLP) and machine learning (ML) increasingly look to incorporate human evaluation in LLM architectures, via techniques like reinforcement learning for human feedback (RLHF) that draw upon HCI methodologies ~\cite{lee2022evaluating, liao2023rethinking, xiao2024human, heuer-buschek-2021-methods}.
On the other hand, researchers across various communities, such as science and technology studies (STS), computer-supported cooperative work (CSCW), and fairness, accountability, and transparency (FAccT) have called for reflection on the potential negative impacts~\cite{smith2022real, hecht2021itstimesomethingmitigating, shen2023shaping},
including a rising chorus of scholars exploring the societal implications of LLM development and the need for responsible AI practices ~\cite{wang2024farsight, do2023that, amershi2019guideline}. 
As various research communities increasingly pursue human-centered methods and questions, there emerges an urgent need for we as the HCI community to scrutinize our own field, and to develop standards for researchers using LLMs and asking HCI-oriented questions.
This work is motivated by the growing need to scrutinize how HCI methodologies are shaping and being shaped by LLM development, ensuring that their influence aligns with scientific rigor and societal benefit.
%

To this end, we contribute a systematic literature review of the 153 LLM-related papers in the last five years of CHI proceedings (2020-2024). 
Our key research questions include: Where have LLMs been applied at CHI? How have researchers used LLMs in their papers? What contributions have LLM-related scholarship made to HCI? What concerns around LLMs do authors articulate?
Our intended audience includes both HCI researchers exploring LLM integration in HCI work and LLM practitioners seeking to understand the current best practices around using LLMs to interact with humans in various domains. 
We identify that LLMs have been taken up in 10 diverse application domains, including Communication \& Writing, Augmenting Capabilities, Education, Responsible Computing, Programming, Reliability \& Validity of LLMs, Well-being \& Health, Design, Accessibility \& Aging, and Creativity. 
Applying \citet{wobbrock2016research}'s framework of research contributions in HCI, we found that LLMs were overwhelmingly used for empirical and artifact contributions, with limited work in theoretical or methodological advances.
To characterize how LLMs are affecting the lifecycles of HCI projects, we identified five roles that LLMs play in research projects: LLMs as system engines, LLMs as research tools, LLMs as participants or users, LLMs as objects of study (e.g., through audits), and users' perceptions of LLMs (e.g., through interview studies of LLM users' experiences). 
We also identified 22 common limitations and risks that authors acknowledged, ranging from qualms around LLMs' performance to concerns around research validity, resource constraints, and potential consequences. 
We found that authors often raise validity and reproducibility concerns around LLM research, despite overwhelmingly studying closed-source LLMs. 

Overall, this work presents an in-depth investigation of the current landscape of how HCI applies and studies LLMs. Towards more rigorous research and responsible design with LLMs, we outline directions for future HCI research at the LLM frontier, and provide actionable recommendations to researchers and practitioners. In summary, we contribute: 

\begin{itemize}[left=0cm] 
    \item a systematic literature review of 153 LLM-related papers from CHI proceedings 2020-2024, resulting in 10 domains where LLMs have been applied, 5 roles that LLMs play in HCI projects, and 29 limitations described by authors;

    \item opportunities for HCI research to leverage LLMs, including under-researched application domains, contribution types, and methodological gaps;

    \item guiding questions for HCI researchers to consider the \textit{validity} and \textit{appropriateness} of a proposed LLM-related study;

    \item an open-source dataset of 153 sampled papers from CHI 2020-2024 with our qualitative codes and paper metadata, publicly available at \url{https://github.com/rrrrrrockpang/llm-chi}.
\end{itemize}

\section{Related Work}

\subsection{\rr{Literature Reviews in HCI}}

HCI has a rich tradition of using systematic \rr{literature} reviews to identify patterns, trends, and limitations of \rr{a research area}~\cite{stefanidi2023literature}. Such reviews provide conceptual frameworks for shared understanding across the field. Many prior works qualitatively analyzed their paper samples to surface high-level themes. For example,  ~\citet{mack2021we} examined 836 accessibility papers over 26 years, coding for common contribution types, communities of focus, and methods. ~\citet{stefanidi2023literature} annotated 189 HCI literature surveys 1982-2022 to \rr{explain} current contributions and topics within HCI. Similarly, ~\citet{dell2016the} manually reviewed 259 HCI4D publications to provide an overview of the space. ~\citet{caine2016local} synthesized standards for sample sizes at CHI by manually extracting data from \rr{each CHI2014 manuscript}. 
Quantitative methods have also been employed to provide broader perspectives on HCI research trends. \citet{liu2014chi} used hierarchical clustering, strategic analysis, and network analysis to map the evolution of major themes in HCI. \citet{cao2023breaking} analyzed patent citations to study the relationship between HCI research and practice. 

\textbf{Our work} builds on this literature to understand LLMs' impact on HCI.
We chose a qualitative approach to provide a deep formative understanding of this rapidly evolving landscape and its impact, not only for HCI researchers, reviewers, and students, but also for researchers in different communities (e.g., AI/NLP) who may be interested in the current state of LLM-ification in HCI, as well as practitioners looking for research-grade guidance on this rapidly evolving space. 

\subsection{\rr{Literature Reviews of LLM Papers}}

Outside of HCI, many fields across computing and social science have used literature reviews to study LLMs' impact on their areas, including reviews of the models, the technical foci, and the societal implications of LLMs. Many of these reviews survey technical advancements, e.g.,  \citet{zhao2023survey} survey methods for training and evaluating core models, \citet{gao2023retrieval} review the state-of-the-art in retrieval-augmented generation, and \citet{guo2024large} review multi-agent approaches. Other efforts have studied the risks posed by LLMs: \citet{weidinger2022taxonomy} taxonomized the harms possible, including discrimination, information hazards, malicious uses, and environmental and economic harms. 

Research has also surveyed trends in how LLMs are being applied in specific disciplines. ~\citet{movva2024topics} collected and analyzed 16,979 LLM-related papers posted to arXiv from 2018 to 2023 to understand trends in LLM research topics. Notably, they found that \textit{society-facing} and \textit{HCI} topics are the two fastest-growing, further showing the urgency of our focus on how the HCI community considers LLM use and implications. \citet{movva2024topics} also found that industry publishes an outsize fraction of top-cited research, but also that industry papers tend to be less open about their models, datasets, and methods. Similarly, ~\citet{fan2023bibliometric} used BERTopic to identify patterns in LLM research 2017-2023. Researchers have additionally employed topic modeling to study LLM usage in fields such as medicine~\cite{barrington2023bibliometric} and education~\cite{lin2024bibliometric}. A recent study shows that papers in behavioral and social science disproportionately favor closed models, despite the availability of powerful, more reproducible open alternatives~\cite{wulff_hussain_mata_2024}. 

\textbf{Our work} focuses on CHI papers, to explore \emph{where} authors applied LLMs in HCI research, and \emph{how} authors leveraged them to make \emph{what} contributions. We extend \citet{movva2024topics}'s quantitative work with an in-depth qualitative analysis of the HCI literature. Our focus on the last five years of CHI papers provides a window into the most recent and most leading-edge work in HCI, since CHI has long been the central and most prestigious venue in the area.

\subsection{How LLMs \rr{Can and Should} Change Research}

There has been substantial debate across the scientific community  on how much LLMs can and should transform research~\cite{birhane2023science}. 
Many papers argue that LLMs are poised to be incorporated in all disciplines, but call for consideration of their limitations. For instance, \citet{aubin2024llm}'s CHI'24 workshop discussed opportunities and responsible integration of LLMs into data work. In computational social science, researchers found that LLMs achieved fair agreement levels with humans on labeling tasks~\cite{ziems2024can}. 
Researchers have also considered whether LLMs can or should influence academic writing~\cite{kim2023chatgpt}. 
A survey of 950,965 papers found a significant increase in the use of LLMs in writing scientific papers, especially in Computer Science~\cite{liang2024mapping}. However, many argue that researchers should ``\textit{avoid overreliance on LLMs and to foster practices of responsible science}~\cite{birhane2023science}.''

\textbf{Our work} extends the discussion on how LLMs are changing and should change research by focusing on the CHI community. We identify the unique roles that LLMs play in HCI research, analyze common limitations reported by authors, and advocate for proactive consideration of these limitations to ensure research rigor.

\section{Methods}

To understand the LLM-ificiation of CHI papers, we performed a literature review of CHI proceedings from 2020-2024. In our study, we focus on generative LLMs, rather than encoder-only models such as BERT or RoBERTa. Via iterative human coding, we assessed (1) the types of contributions common in LLM-focused HCI scholarship, (2) the roles that LLMs are playing in research projects; and (3) the limitations that researchers are disclosing in their papers.

\subsection{Data}

We first gathered the full-text proceedings of CHI from 2020-2024, which at the time of writing represented the most recent five years of cutting-edge HCI research. In 2020, OpenAI's GPT-3 was released, marking a leap in language models' predictive capabilities. LLMs then became more accessible to researchers through APIs and open-sourced models.\footnote{https://techcrunch.com/2020/06/11/openai-makes-an-all-purpose-api-for-its-text-based-ai-capabilities/}
We chose CHI for two reasons. 
First, CHI is the flagship international conference on HCI. All papers undergo rigorous peer review, and publications have  significant impact on HCI research generally. Similar prior literature reviews chose CHI as a representative sample to identify trends in HCI~\cite{linxen2021weird, bartneck2009scientometric}. 
Second, CHI papers span a wide range of application areas and methodologies, (e.g., CHI 2024 had 16 subcommittees) giving this work broad representation. We acknowledge that ACM SIGCHI sponsors 26 HCI conferences,~\footnote{https://sigchi.org/conferences/} which have more focused scopes. 
Our sample should be considered \emph{generative} rather than exhaustive, and our work can spur future analysis of more focused conferences. 

\begin{figure}
    \centering
    \includegraphics[width=0.95\linewidth]{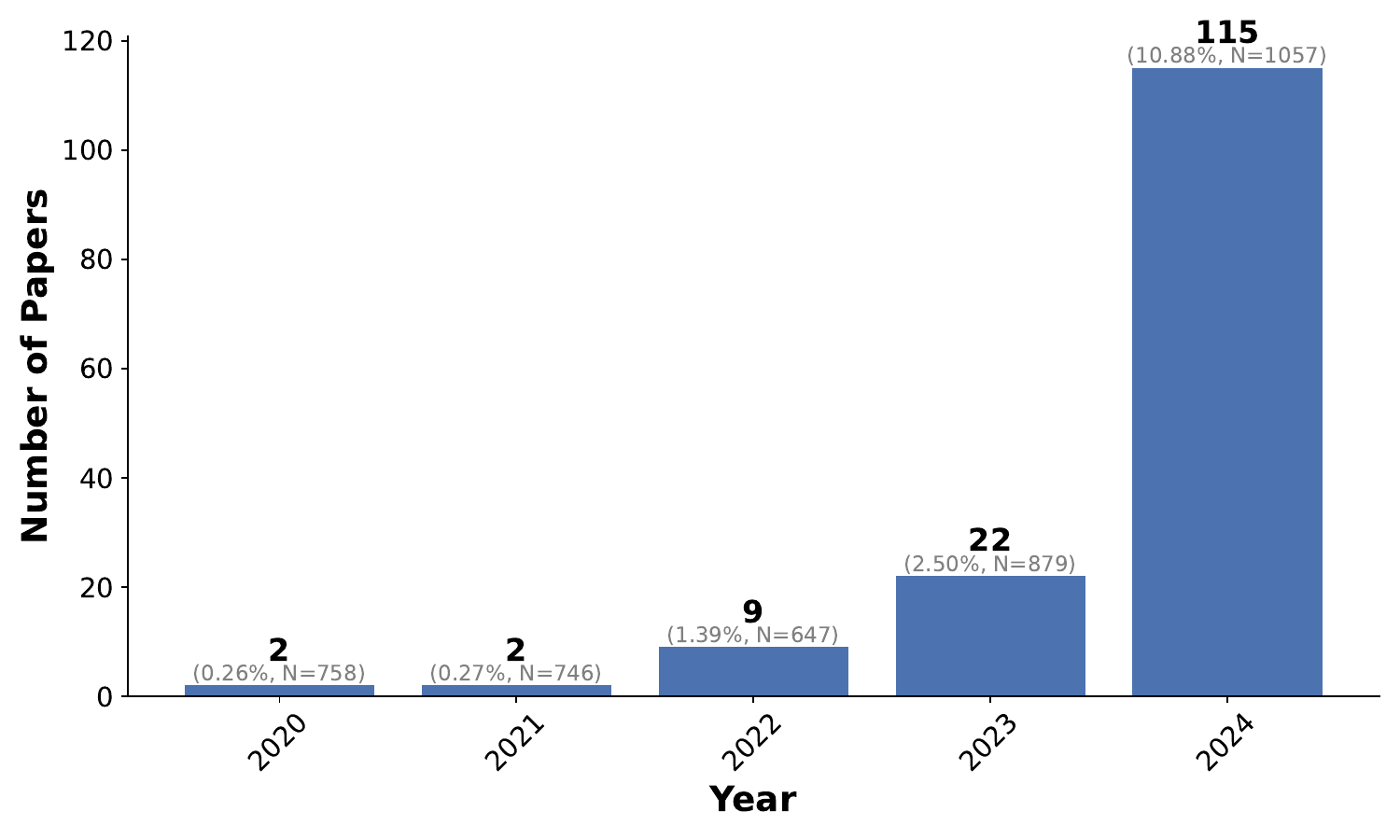}
    \caption{The raw number of LLM-related papers, followed by the percentage of the total number of papers in each year 2020-2024.}
    \Description{A bar chart that shows the number of LLM-related CHI papers in our sample across year 2020-2024. There are 2 papers in 2020, 2 in 2021, 9 in 2022, 22 in 2023, 115 in 2024.}
    \label{fig:paper_distribution}
\end{figure}

\begin{figure}
    \centering
    \includegraphics[width=0.6\linewidth]{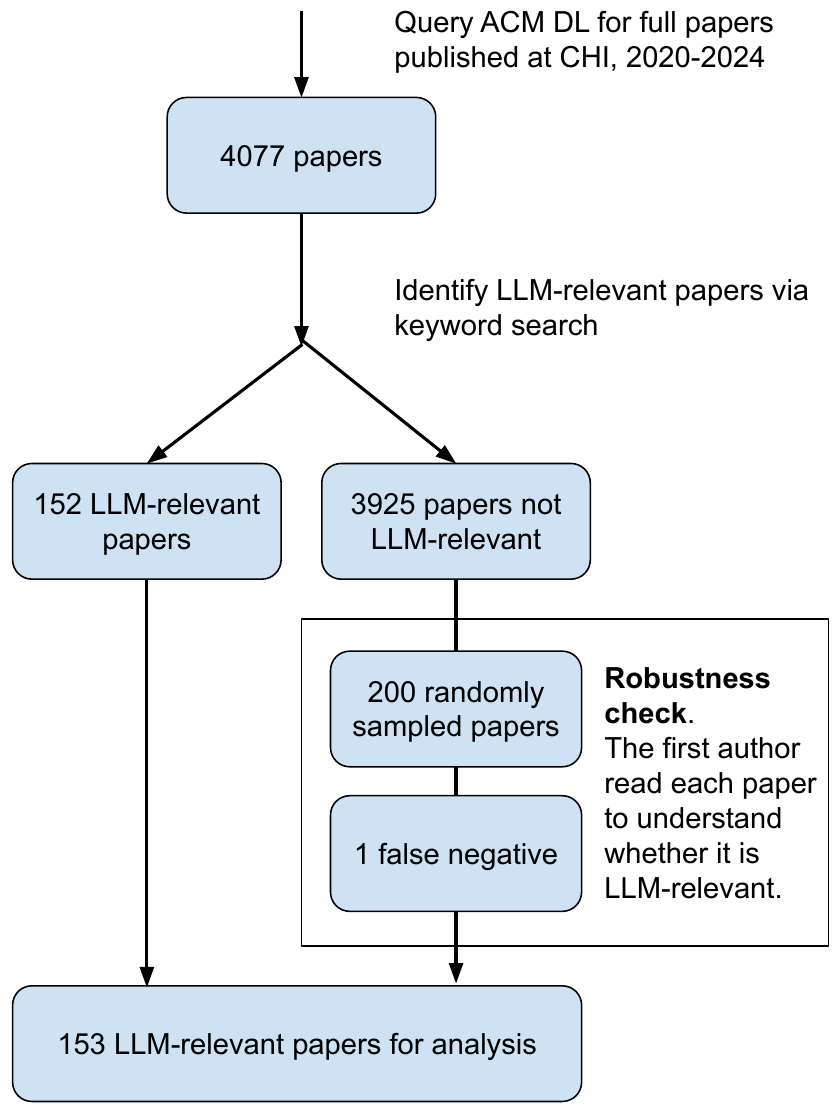}
    \caption{A flow diagram on our sample selection and refinement process.}
    \Description{A flow chart diagram of our sample selection and refinement process. On the top is an arrow pointing to 4077 papers, which is the overall retrieved sample. On the arrow, there is text that shows "Query ACM DL for full papers published at CHI2020-2024." Then this box points to two other boxes. The first one reads "152 LLM-relevant papers," the other "3925 not LLM-relevant." The second points to a rectangle that reads "Robustness check. The first author read each paper to understand whether it is LLM-relevant." Under this rectangle shows 1 false negative identified from 200 randomly sampled papers. Together, we have 153 LLM-relevant papers for analysis.}        \label{fig:analysis_workflow}
\end{figure}

\rr{Our process follows an adapted PRISMA statement~\cite{moher2010preferred, page2021prisma} and is summarized in \autoref{fig:analysis_workflow}.} We began by assembling a corpus of CHI papers, and filtering for LLM-relevant works. We contacted ACM Digital Library (DL) 
in July 2024 for full papers (excluding extended abstracts, doctoral consortium submissions, and other non-paper artifacts) published at the conferences from 2020-24.
This search resulted in an initial corpus of 4,077 papers.
We then filtered each paper's title and abstract by a set of keywords: \texttt{``language model'', ``llm'', ``foundation model'', ``foundational model'', ``GPT'', ``ChatGPT'', ``Claude'', ``Gemini'', ``Falcon''}. 
This filter resulted in a corpus of 152 LLM-relevant papers. \autoref{fig:paper_distribution} shows the breakdown of papers, as well as the percentage of the paper numbers in each year. 
We did not search full text on the CHI proceedings because our early investigation showed that it resulted in substantially more false positives (e.g., a paper might have one sentence that mentions their implications ``\textit{in the age of LLMs}''). We also did not include general keywords (e.g., ``artificial intelligence''), since our focus is papers that include LLMs, rather than capturing the wider field of AI research, as has been studied in prior work on the human-AI interaction literature~\cite{yang2018mapping, grudin2009ai, amershi2019guideline, yang2020re}.

We additionally ensured the robustness of our filtering procedure by validating our corpus against false negatives.
We conducted a stratified sampling of 200 papers that were initially found to be \textit{not} LLM-relevant.
The first author then read each of the 200 papers to check whether the work was LLM-relevant.
Our review found one paper (N=1, 0.5\%) that was not identified by our keyword search procedure. 
This paper mentioned GPT-4 just once, in their method section, without mentioning any other keywords in our list. We added this paper to our final corpus (N=153, see  \autoref{fig:analysis_workflow}).

\subsection{Analysis}

To analyze the 153 papers, we applied an iterative process to develop a codebook. The initial codebook included deductive codes based on our four research questions. 
For two of our research questions, we used existing taxonomies to seed our codebooks: 
on the contribution type, we used \citet{wobbrock2016research}'s taxonomy of research contributions in HCI, and 
on the application domains for each paper, we used a taxonomy from \citet{stefanidi2023literature}.
For the rest of the questions---on the roles of the LLMs in each paper, the limitations and risks of the research---we generated initial codebooks during the iterative open coding process. 

We conducted four iterations of independently applying and updating the codebook, using a randomly selected set of 10 papers for each iteration. After each set, the research team came together to refine or merge existing codes, add new codes, and resolve disagreement through consensus.
Throughout, we computed interrater reliability (IRR) using Krippendorff's alpha to guide our discussions.\footnote{Note that Fleiss’ kappa can also be applied to the IRR analysis of the complete nominal data in our case. We chose Krippendorff's alpha in line with prior literature review~\cite{mack2021we}.} The final alpha values are $\alpha_{\text{Contribution Types}} = 0.866$, $\alpha_{\text{Application Domains}} = 0.849$, $\alpha_{\text{LLM roles}} = 0.773$, $\alpha_{\text{Limitations \& Risks}} = 0.887$\footnote{The $\alpha$ over our initial 29 low-level code is $0.633$}. All values are comparable to prior work~\cite{mack2021we, lee2024design}.
This process led to a codebook of 51 low-level codes (see Supplementary Materials for the process and the codebook). 
Finally, the remaining papers---those that had not been used for codebook development---were split into three sets and coded independently by the three researchers who all participated in the codebook development. During this step, the authors met regularly to discuss any emergent concerns, and disagreements were resolved through consensus. 

\subsection{Research Positionality}

We acknowledge that our academic and professional backgrounds have shaped our perspectives on this topic. All authors had experience using LLMs directly or studying users' perceptions of LLM-powered systems, and had experience working in responsible computing. The authors' expertise covers fields including HCI, NLP, computational social science, accessibility, machine learning, fairness, sociotechnical systems, and usable security and privacy. Collectively, we are US-based researchers at three different R1 universities and one US-based research institute. 
\begin{table*}
\small
\setlength{\tabcolsep}{5pt}
\renewcommand{\arraystretch}{1.3}
\begin{tabular}{c>{\raggedright\arraybackslash}p{4cm}>{\raggedright\arraybackslash}p{12cm}}
\toprule
\hline
& \textbf{Code} & \textbf{Definition} \\
\midrule
\multirow{10}{*}{\rotatebox[origin=c]{90}{\adtext{\textbf{Application Domains}}}} &
\multicolumn{2}{l}{\textit{Where have LLMs been applied in CHI papers?}} \\
\cmidrule(l){2-3}
& \ad{Communication \& Writing} & On various writing and communication tasks, which often target writers as the primary user groups. \\
& \ad{Augmenting Capabilities} & On technologies to enhance human performance and productivity, often in the physical world. \\
& \ad{Education} & On learning experiences for students and pedagogical methods for educators. \\
& \ad{Responsible Computing} &  On ethical and societal implications of computational systems, particularly in high-stakes domains. \\ 
& \ad{Programming} & On various aspects of software development and programming tasks. \\
& \ad{Reliability \& Validity of LLMs} & On evaluating and improving LLM outputs themselves. \\
& \ad{Well-being \& Health} & On managing health-related disorders/illnesses, or interactions with health data or healthcare providers. \\
& \ad{Design} & On various types of design work, which often target designers as the primary user group. \\
& \ad{Accessibility \& Aging} & On population with disabilities and older adults. \\
& \ad{Creativity} & On the creativity process and creativity support tools, which often overlaps with other domains. \\
\\
\midrule
\multirow{6}{*}{\rotatebox[origin=c]{90}{\roletext{\textbf{LLM Roles}}}} &
\multicolumn{2}{l}{\textit{How do CHI papers leverage these models?}} \\
\cmidrule(l){2-3}
& \role{LLMs as system engines} & LLMs function as core elements within systems, prototypes, algorithms, and programming frameworks. \\
& \role{LLMs as research tools} & LLMs perform research tasks traditionally executed by researchers in a research project, such as data collection, analysis, and writing. \\
& \role{LLMs as participants \& users} & LLMs simulate human responses and behaviors, or act as users or participants in an interaction. \\
& \role{LLMs as objects of study} & LLMs' inner mechanism, properties, performance are evaluated. \\
& \role{Users' percetions of LLMs} & LLMs or tools (e.g., ChatGPT) are studied to understand user perceptions in different contexts. \\
\midrule
\multirow{8}{*}{\rotatebox[origin=c]{90}{\limittext{\textbf{Limitations \& Risks}}}} &
\multicolumn{2}{l}{\textit{What are the concerns by the authors at CHI?}} \\
\cmidrule(l){2-3}
& \makecell[l]{\limit{Limitations on}\\\limit{LLM Performance}} & Limitations specifically on the LLM capability to output the desired output. This includes \emph{LLM bias toward different groups}, \emph{limited data coverage in the training data}, \emph{non-deterministic response}, \emph{hallucination}, \emph{unspecific errors and biases} \\
& \makecell[l]{\limit{Limitations on}\\\limit{Research Validity}} & Limitations to the extent which an instrument measures what it claims to measure in the paper. This includes internal and/or external validity across users, contexts, models, and prompts. \\
& \limit{Limitations on Resource} & Limitations on computational and financial resources to open or closed source LLMs. This includes \emph{computational cost}, \emph{financial cost}, \emph{lack of evaluation standards} \\
& \limit{Risks to Society} & Potential negative and long-term outcomes, risks, or unintended effects may arise from the artifact or study. This includes \emph{economic harms}, \emph{representational harms}, \emph{misinformation harms}, \emph{malicious use}, \emph{hate speech}, and \emph{environmental harms}. \\

\hline
\bottomrule
\end{tabular}
\caption{Domains where LLM applications are developed, roles of LLMs in HCI projects, and acknowledged risks and limitations. Note that we did not include contribution types in this table. A paper can have \textit{multiple} (sub-)codes.}
\label{tab:code}
\end{table*}

\section{Results}
 
Our analysis reveals \emph{where} LLMs have been applied at CHI, \emph{how} researchers have leveraged these models, and \emph{what} contributions they made to the field of HCI.
In parallel, we taxonomize the common \emph{limitations and risks} articulated by authors (see \autoref{tab:code}).

\subsection{\adtext{Application Domains}}

We found 10 diverse domains in which HCI researchers have applied LLMs (\autoref{tab:code}). We elaborate each in this section.

\ad{Communication \& Writing} (22.88\%, N=35): This domain emerges as the most-studied area, spanning both specific writing tasks and \textit{AI-mediated communication (AIMC)} \cite{hancock2020aimc}, in which intelligent agents modify, augment, or generate messages to achieve communication goals.
Many of these works imagine writers as the target LLM user, in tasks from personal diaries~\cite{kim2024diarymate} and email composition~\cite{buschek2021impact}
to storytelling~\cite{chung2022talebrush}, screenplay creation~\cite{mirowski2023co}, and general creative writing~\cite{chakrabarty2024art, wang2023popblends}. 
For instance, researchers have examined writers' attitudes toward collaborating with LLMs~\cite{li2024value}, including how writers choose prompting strategies~\cite{dang2023choice} and users' perception of AIMC support in a variety of writing tasks, such as idea generation, translation, and proofreading \cite{fu2024from}. 
Researchers have also examined how LLMs might introduce implicit bias to the writing process~\cite{fu2023comparing, jakesch2023co}.

\ad{Augmenting Capabilities} (16.99\%, N=26): This domain includes papers that develop technologies to enhance human \emph{performance} and \emph{productivity} by altering how we engage with technology and information. Some attempt to bridge the physical and digital worlds in scenarios such as video conferencing~\cite{liu2023visual} and mixed reality~\cite{de2024llmr}.
Many also study the future of work and productivity. \citet{fok2024marco} leverages LLMs to support sensemaking on business document collections, while \citet{kobiella2024if} studied how ChatGPT usage affects professionals' perceptions of workplace productivity. Several papers also developed tools to enhance productivity in academic research, building new approaches for sensemaking of literature~\cite{lee2024paperweaver} and research idea generation~\cite{wang2024evaluating}.

\ad{Education} (14.38\%, N=22): This domain explores the potential of LLMs to enhance learning experiences for students and improve pedagogical methods for educators. 
For students, research examined learners' existing interactions with LLMs, including \citet{belghith2024testing}'s investigation of middle schoolers' approaches to and conceptions of ChatGPT. Several works explored using LLMs as learning aids in specific subject areas, such as math~\cite{zhang2024mathemyths}, vocabulary acquisition~\cite{lee2024open}, and programming~\cite{jin2024teach}.
For educators, studies examined the LLMs' integration into teaching. \citet{han2024teachers} found that teachers are excited about potential benefits, namely LLMs' ability to generate teaching materials and provide personalized feedback to students; however, teachers and parents are both concerned about their impact on students' agency in learning, and potential exposure to bias and misinformation. Researchers have also designed LLM-based tools to assist teachers in domains such as cyberbullying education~\cite{hedderich2024piece} and environmental science instruction~\cite{cheng2024scientific}.

\ad{Responsible Computing} (12.42\%, N=19): This explores ethical and societal implications of computing systems, particularly in high-stakes domains and for vulnerable populations. It touches on issues like fairness, information hazards, and privacy.
\rr{Several} studies have examined how marginalized groups perceive LLMs, focusing on gender~\cite{sun2022pretty, ma2024evaluating}, religion~\cite{precel2024canary}, and other intersectionalities~\cite{fujii2024silver}. 
Research also identified the risks LLMs pose to to those seeking information online. For instance, \citet{sharma2024generative} investigated how LLM-powered search systems might amplify echo chambers, 
while \citet{oak2024understanding} \rr{studied} the use of LLMs by underground incentivized review services. \citet{zhou2023synthetic} outlined approaches to addressing LLM-generated misinformation.
Papers also addressed a range of privacy issues, including online surveillance on social media~\cite{choksi2024under}, users' navigation of disclosure risks and benefits when using LLM-based conversational agents~\cite{zhang2024s}, and general privacy knowledge~\cite{chen2024an}.
Finally, we identified papers that integrate LLMs into interactive tools designed to facilitate responsible computing practices~\cite{wang2024farsight, pang2024blip}. 

\ad{Programming} (11.11\%, N=17): This domain automates and improves software development and programming tasks, including papers related to data science, analytics, and visualization systems. 
Many papers develop tools to facilitate code creation. For instance, \citet{liu2023wants} introduced a novel method called grounded abstraction matching, powered by Codex, to assist non-expert programmers in guiding code generation. Other tools support programmers by providing no-code platforms for traditionally complex programming languages~\cite{kim2022stylette}, explaining code generation~\cite{yan2024ivie}, and aiding in programming language learning~\cite{chen2024learning}.
We also include work on ``\textit{prompt engineering}'' in this category, such as prompt sharing~\cite{feng2024coprompt}, direct manipulation of LLM outputs~\cite{masson2024directgpt}, and visual prompt comparison~\cite{arawjo2024chainforge}. These studies used programming tasks for evaluation, reflecting the 
broader trend of incorporating prompt engineering into software engineering~\cite{White2024}.
On a critical note, \citet{kabir2024samia} analyzed ChatGPT’s responses to 517 StackOverflow programming questions, revealing that 52\% of the answers contained incorrect information and 77\% were verbose.

\ad{Reliability \& Validity of LLMs} (10.46\%, N=16): This domain focuses on evaluating and improving LLM outputs. 
The first stream of work includes analyses determining the validity of applying LLMs to specific contexts. For example, \citet{he2024if} compared GPT-4 and Mechanical Turk pipelines for sentence labeling tasks from scholarly articles, showing that combining crowd and GPT-4 labeling increases accuracy. Another example is \citet{kabir2024samia},  evaluating the validity of using LLMs' to answer programming questions. 
The second stream involves tools designed to enhance the reliability or validity of LLM outputs. For instance, HILL identifies and highlights hallucinations in LLM responses, allowing users to handle responses with greater caution~\cite{leiser2024hill}. EvalLM enables interactive evaluation of LLM outputs based on user-defined criteria across multiple prompts~\cite{kim2024evallm}. AI Chain is a visual programming tool for crafting LLM prompts, which improved the quality of task outcomes as well as the transparency, controllability, and the sense of collaboration when interacting with the black-box LLMs~\cite{aichain2022wu}.

\ad{Well-being \& Health} (9.15\%, N=14): This domain refers to the management and prevention of health-related disorders and illnesses, or interactions with health data or with healthcare providers.\footnote{While some health-related conditions may fall under accessibility, such as chronic illness~\cite{good2014accessing}, we decide according to how the condition was treated: papers that adopt a social model of the condition or disability (i.e. that the incompatible design of society with the person's condition is the ``problem'') are \emph{Accessibility}, and those that adopt a medical model (i.e. that the person's condition is the ``problem'') are classified here under Well-being \& Health~\cite{haegele2016disability}.} 
One thread of work involves assisting practitioners in providing better care. For example, \citet{yang2023harness} designed a GPT-3-based decision support tool that draws on the biomedical literature to generate AI suggestions. \citet{yildirim2024multimodal} worked with radiologists to explore the design space for incorporating LLMs into radiology.
Another thread involves support for patients in self-tracking, self-diagnosing, and self-managing their illnesses. For instance, \citet{sharma2024facilitating} used a fine-tuned GPT-3 model to improve self-guided mental health interventions through cognitive restructuring, a technique to overcome negative thinking. MindfulDiary leveraged GPT-4 to support psychiatric patients’ journaling~\cite{kim2024mindfuldiary}. \citet{stromel2024narrating} found that GPT-generated data description can effectively complement numeric fitness data.

\ad{Design} (8.50\%, N=13): This domain captures papers whose target audience is designers. For example, HCI researchers have produced LLM-powered tools that facilitate the design process for practitioners, such as mobile UI design~\cite{huang2024automatic, xiang2024simuser, duan2024generating}, landscape design~\cite{huang2024plantography}, interior color design~\cite{hou2024c2ideas}, and multimodal application design~\cite{lin2024jigsaw}. On the other hand, \citet{liao2023designer} interviewed 23 UX practitioners to explore the design space around LLMs supporting ideation, including their needs around model transparency.

\ad{Accessibility \& Aging} (7.84\%, N=12): This domain focuses on people with disabilities and older adults. 
We found diverse accessibility contexts, including the blind or low-vision (BLV) community~\cite{zhang2024designing, taeb2024axnav}, people with autism~\cite{jang2024s}, learning disabilities involving Augmentative and Alternative Communication (AAC)~\cite{valencia2023less}, and situational impairments~\cite{liu2024human}, as well as papers on older adults~\cite{xygkou2024mindtalker}. However, we did not find papers on the deaf or hard of hearing (DHH) community or motor or physical impairments, which are generally the second and third most prevalent in terms of accessibility paper counts~\cite{mack2021we}.

\ad{Creativity} (5.88\%, N=9):
This domain covers the creative process and creativity support tools. \citet{chakrabarty2024art} proposed the Torrance Test of Creative Writing (TTCW) to directly scrutinize whether LLMs are ``creative'' via a story writing task. Similarly, Jigsaw presented a creativity support tool to assist \emph{designers} with prototyping multimodal applications by chaining multiple generative models~\cite{lin2024jigsaw}.

\subsection{\cttext{Contribution Types}}

The above application domains were primarily addressed through (1) \ct{empirical contributions} (98.70\%, N=151)---often to understand a population’s view toward or use of LLMs or specific LLM-powered tools---and (2) \ct{artifact contributions} (61.44\%, N=94), which involve building a tool. These two contribution types frequently occur in combination, in studies where authors first build an artifact and then empirically test it with users.
For artifact contributions, we observed that LLM-powered systems have a wide range of fidelity levels, from fully open-sourced systems with GitHub repositories to simple wireframes. The dominance of LLMs in these systems also varied, with some systems using LLMs throughout the entire pipeline and others using them only for processing textual data. We applied the code ``\textit{artifact contribution}'' to a paper when authors claimed that LLMs are (or would be) a part of the system. The high frequency of artifact contribution (61.44\% in our sample in contrast to 24.50\% at CHI overall~\cite{wobbrock2016research}) may indicate that LLMs might have lowered the barrier to prototype research artifact of high quality, a point we unpack further in \ref{sec:dis-hci-prototyping}.

The remaining five contribution types occurred less frequently, with one \ct{survey contribution} and no \ct{opinion} \ct{contributions}. Distinguishing between methodological and artifact contributions can be challenging, as some methods are embedded in a system. Per ~\cite{wobbrock2016research}, we used \ct{methodological contribution} to refer to research method contributions in HCI. Methods for creating multimodal mobile applications, for example, were not included. Overall, we found 16 (10.46\%) methodological contributions, such as LLM-augmented methods to enhance UX evaluation~\cite{kuang2024enhancing}, generate synthetic user data~\cite{wang2024evaluating}, and provide metrics to measure creativity in LLMs~\cite{chakrabarty2024art}.  
We found 8 \ct{theoretical contributions} (5.23\%), ranging from a framework for collaborative group-AI brain-writing~\cite{shaer2024ai}, a conceptual framework to bridge the gulf of envisioning~\cite{subramonyam2024bridging}, and a design space for intelligent writing assistants~\cite{lee2024design} (also a \ct{metareview} \ct{contribution}).
\ct{Dataset contributions} were less common (N=6, 4.0\%). In the \roletext{LLM roles} section, several papers used LLMs to generate synthetic datasets, which may lower the barrier to creating large, diverse datasets for thorough evaluation, yet curating benchmark datasets of real users that can test the performance of LLMs at scale remains a challenge~\cite{lee2022coauthor}.

\subsection{\roletext{LLM Roles}}

We identified five roles that LLMs play in HCI research (\autoref{fig:llm-role}). While \autoref{fig:llm-role} may not fit every project given the interdisciplinary nature of our field, it reflects our sample, which primarily offers empirical contributions.

\role{LLMs as system engines} (62.74\%, N=96):  In this role,
LLMs function as core elements within systems, prototypes, algorithms, and programming frameworks. 
One way LLMs can be used in systems is to \emph{generate content}, e.g., ideas, code, and conversations.  
For example, Farsight used LLMs to generate ideas to identify potential harms of AI applications~\cite{wang2024farsight}, and GenLine used LaMDA to generate code from users' natural language~\cite{jiang2022discovering}. 
MindTalker, a GPT-4 conversational agent, supports people with early-stage dementia by reducing loneliness~\cite{xygkou2024mindtalker}. 
On the other hand, LLMs may be used to \emph{process information and extract insights}, e.g., by retrieving or summarizing from large, unstructured datasets. For instance, PaperWeaver deduces users' research interests from their paper collection on Semantic Scholar~\cite{lee2024paperweaver}, while Memoro interprets user needs from the users' conversation history~\cite{zulfikar2024memoro}. Visual Captions employed a fine-tuned LLM to predict user intent using the sentences in a video conferencing call~\cite{liu2023visual}. 
Systems integrate LLMs at different levels. Some systems' main functions rely on a carefully-designed system prompt, often in a form instructions to a conversational agent~\cite{tang2024emoeden}, while others used LLMs as one~\cite{hao2024advancing} or more~\cite{lam2024concept} step(s) in a complex pipeline. 
On another axis, the LLM-powered tools can range from a fully-functioned open-source system~\cite{wang2024farsight} to design prototypes that elicit important empirical insights~\cite{yang2023harness}.

\definecolor{customgreen}{HTML}{13ae5c}  
\definecolor{customyellow}{HTML}{ffa629}   
\definecolor{custompurple}{HTML}{9647fc}  
\newcommand{\coloredarrow}[1]{\textcolor{#1}{$\boldsymbol{\xrightarrow{\ \ }}$}}
\begin{figure*}
    \centering
    \includegraphics[width=\linewidth]{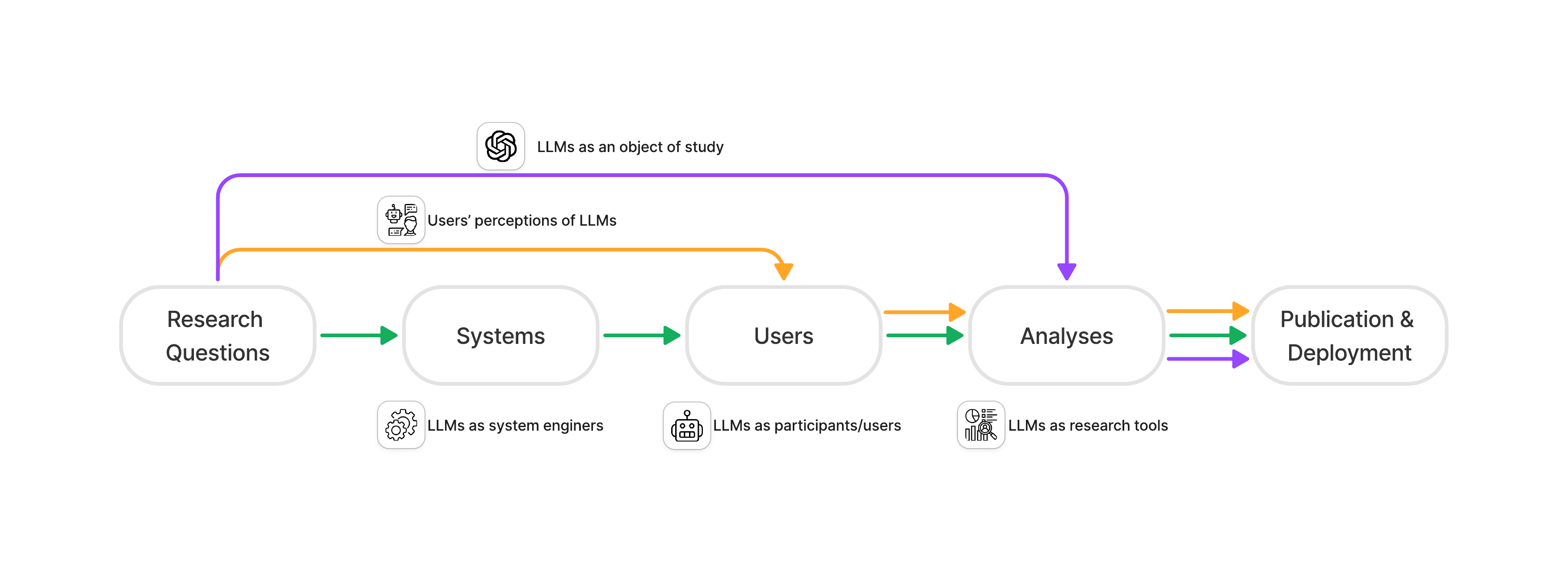}
    \caption{Overview of roles that LLMs can play throughout common HCI research stages. LLM roles are depicted with icons and text, and arrows represent common empirical studies: \protect\coloredarrow{customgreen} system building studies; \protect\coloredarrow{customyellow} user studies; \protect\coloredarrow{custompurple} data science studies.}
    \Description{A flowchart that shows an overview of different roles that LLMs can play throughout common HCI research stages. The stages are Research Questions, Systems, Users, Analyses, Publication \& Deployment. There are three colored arrows: the first is green, and it connects from left to right all the five stages. This denotes system-building studies. The second is an orange arrow that connects the Research Question, Users, Analyses, and Publication \& Deployment. This indicates user studies without building systems. The third is a purple arrow that connects Research Questions, Analyses, and Publication \& Deployment. This indicates data science studies.}
    \label{fig:llm-role}
\end{figure*}

\role{LLMs as research tools} (9.15\%, N=15): 
We found several authors used LLMs to perform tasks traditionally executed by researchers or research assistants, including \emph{data collection, analysis, or writing}. For example, \citet{choksi2024under} applied LLMs to conduct qualitative coding on social media posts on NextDoor. They first developed a codebook, manually labeled 340 posts, and then adjusted the codebook prompts before using GPT-4 to tag the rest of the posts from the sample. 
Such LLM-augmented workflows were often also claimed as a methodological contribution, or packaged as a system that other researchers could use. For example, \citet{wang2024human} introduced a multi-step human-LLM collaborative method for qualitative coding. In this process, LLMs generate labels and explanations, a verifier model assesses the quality, and humans re-annotate the subset of low-quality labels. \citet{ding2024leveraging} contributed a LLM-based method to identify critical online discussion patterns at scale to inform national health outcomes.
Similarly, \citet{lam2024concept} proposed LLooM, a LLM-powered Python package to iteratively synethesize concepts over a sample of text. 

Another thread involved using LLMs to \emph{generate data} for research purposes. For example, \citet{sun2022pretty} used GPT-2 to generate a corpus of 96,600 artificial greeting messages to study gender bias in greeting card messages and facilitate future research on this topic. \citet{ko2024natural} introduced a LLM-based framework that takes Vega-Lite specification as input to generate diverse natural language datasets, such as captions, utterances, and questions about the visualization. \citet{feng2024mud} uses LLMs to automatically mine UI data from Android apps. 
These papers often generate synthetic datasets and conduct analyses as part of their contributions.

\emph{Additional Analysis}: As using LLMs to perform research tasks is becoming new research methodology~\cite{aubin2024llm}, we examined authors' justification of this role. For all 15 papers, authors justified their LLM usage, explained LLMs' capabilities and suitability for the task, and cited relevant prior work. All but one paper provided further experimental validation. These validations took the form of comparison user studies with non-LLM baselines, manual validation of system outputs, and human or computational quality assessments. All but one paper relied on humans for the evaluation, and this exception \cite{sun2022pretty} used computational methods for quality analysis.

\role{LLMs as participants \& users} (7.19\%, N=11):
This category uses LLMs to simulate human responses and behaviors, acting as users or participants. One line of work relied on the assumption that LLMs can create believable proxies for human behaviors, known as ``\textit{personas}''. 
Personas were proposed decades ago in HCI ~\cite{pruitt2003personas} to guide user research, by creating abstract representations of users that provide valuable insights into user needs, behaviors, and preferences. 
By prompting LLMs to create such personas, researchers aim to approximate user feedback.
For example, Impressona generated on-demand feedback from writer-defined AI personas of their target audience~\cite{benharrak2024writer}.
Similarly, \citet{hedderich2024piece} built Co-Pilot for teachers to prepare them to chat with students about cyberbullying.
Another thread includes works on using LLMs to simulate potential user feedback for systems or designs. \citet{duan2024generating} applied GPT-4 to automate heuristic evaluation
via a Figma plugin. Likewise, SimUser leveraged LLMs to simulate usability feedback~\cite{xiang2024simuser}. \citet{hamalainen2023evaluating} explicitly studied the validity of using LLMs to generate synthetic user research data, concluding that GPT-3 can generate believable answers to open-ended questionnaires about experiencing video games as art.

\emph{Additional Analysis}: Using LLMs as participants \& users constitutes a novel research methodology, so we also analyzed authors' justification of their methodology. We found that 10/11 papers provided both textual justification and experimental validation. Similar to the LLM-as-research-tools papers, the text justifications are also supported by citations to relevant prior work in NLP. 
Experiments similarly spanned user studies, as well as human and computational analysis. Rather than justifying the usage, \citet{cuadra2024illusion} studied this very topic with a more critical lens and demonstrated the validity concerns inherent to LLM use in chatbots as humans, which \citet{wang2024large} and \citet{agnew2024illusion} address from an ethical perspective as well.

\role{LLMs as objects of study} (9.80\%, N=15): 
This category contains papers that explore LLMs' underlying mechanisms and properties,
including training datasets, response outputs, and issues (e.g., hallucination). Some works study potential problems inherent to LLMs themselves. For instance, \citet{precel2024canary} scrutinized common LLM training datasets and found that a disproportionate amount of content authored by Jewish Americans is used for training without their consent, and \citet{sun2022pretty} studied the gender bias of GPT-2 generated text. 
Other works focus on the ecological validity of applying LLMs in a particular context.
\citet{kabir2024samia} conducted an empirical study of the characteristics of ChatGPT answers to StackOverflow questions, evaluating whether LLMs are appropriate to use in the context of Q\&A for programming questions.  
Several studies examine the validity of using LLMs for crowdsourcing tasks~\cite{he2024if}. 

\role{Users' Perceptions of LLMs} (23.53\%, 36):
This category includes studies on how users perceive LLMs or LLM-powered tools. Papers often examine a particular population's perception and usage of a public LLM-powered tools (e.g., ChatGPT) to  create a design space or surface challenges and opportunities. For fine-grained insights, we exclude user studies that evaluate a system where the research artifact is the main contribution.
For example, papers have studied how LGBTQ+ individuals experience using chatbots for mental health support~\cite{ma2024evaluating}. Using the case of CareCall---a deployed chatbot for socially isolated individuals in South Korea---\citet{jo2023understanding} attempted to understand how LLM-driven chatbots can support public interventions. 
Other works have studied how diverse users perceive and interact with LLMs or LLM-powered chatbots, including teachers~\cite{hedderich2024piece, tan2024more}, middle schoolers~\cite{belghith2024testing}, creative writers~\cite{gero2023social}, and performance artists~\cite{jones2023embodying}.
Several works have also examined LLMs’ effects on users. For example, \citet{wester2024as} studied how LLMs deny user requests, and \citet{jakesch2023co} examined how users write social media posts with LLM assistance.

\subsection{\limittext{Limitations}}

This section covers four top-level codes and 22 main sub-level codes for the limitations and risks discussed in our sample. Coding the limitations is not a trivial task, as not every paper has a dedicated ``limitations'' section. We found 94.77\% (N=145) papers with a dedicated section for limitations (i.e., with ``limitations'' in the section title) and 14.38\% (N=22) papers with a dedicated ethics or impact statement. Our analysis was primarily based on the limitations section; if there was no limitations section, we read through the paper to find potential mentions of limitations.

\subsubsection{\limittext{LLM Performance} (42.48\%, N=65)} 
The top-level code refers to limitations on LLMs' capability to generate the desired output. These limitations highlight areas where the LLM’s performance may fall short of expectations.  

\limit{LLM bias toward different groups} (11.11\%, N=17): This limitation recognizes that LLMs' disparate representation across different populations. For example, \citet{shin2024paper} noted the GPT-3 and DALLE-2 in their system might output and perpetuate gender and racial stereotypes, including a higher chance of featuring white men rather than users in other racial groups. This limitation also includes cases where LLMs \emph{fail to} model certain user groups---the absence of those users. \citet{ma2024evaluating} stated that LLM-based chatbots failed to ``\textit{recognize complex and nuanced LGBTQ+ identities and experiences, rendering the chatbots' suggestions generic and emotionally disengaged.}''

\limit{Limited data coverage in the training data} (9.80\%, N=15): Authors explicitly mentioned that LLMs' training data might be insufficient or outdated. For instance, \citet{lee2024open} found they needed extra engineering steps to use an LLM with their Korean-speaking participants, which they attributed to ``\textit{GPT-4's underperformance in non-English languages}''. When prompting LLM conversational agents to display empathy using elicitations from Reddit, \citet{cuadra2024illusion} acknowledged that they are not aware of the distribution of the training data, and are therefore unable to tell whether the data used in the study has been covered by GPT-4. 

\limit{Non-deterministic response} (7.84\%, N=12): Authors often recognized that LLM responses are probabilistic, and could change unpredictably even when given the same prompt. \citet{gu2024analysts} recognized that their LLM's explanations were not fully controlled, because they used real-time responses from commercial models. \citet{chen2024an} attributed the inconsistency of generated data to the ``\textit{inherent randomness embedded in the output of LLMs}.'' This, however, can be alleviated by changing the sampling temperature to zero~\cite{ouyang2023llm} or using guided generation~\cite{zekun2023guiding}.

\limit{Hallucination} (8.50\%, N=13): LLMs can produce inaccurate or entirely fabricated information. 
\citet{hoque2024the} explicitly pointed out that ``\textit{LLMs can generate hallucinations},'' which may ``\textit{alter the dynamic for such authors [in their study] when using an LLMs}'' but later stated that studying the effect is out of their study scope.
Though Retrieval Augmented Generation (RAG) systems may help alleviate this problem in the future \cite{izacard2021leveraging, gao2023retrieval}, applications that leverage this approach can still suffer from hallucination issues. For instance, \citet{zulfikar2024memoro} stated that using LLMs ``\textit{in information retrieval can lead to hallucinated answers that do not exist in the dataset}.'' 
To ensure validity, works such as PaperWeaver~\cite{lee2024paperweaver} attempted to evaluate the system's performance against hallucination by collecting annotations of factual correctness for 60 descriptions in their study, but not all papers grappled with this problem as explicitly.

\limit{Unspecified errors and biases} (16.99\%, N=26): 
This the most common code related to LLM performance. Papers vaguely recognize the problems of LLMs, but authors did not specify the exact errors due to the models' black-box nature. For example, \citet{li2024omniactions} stated that ``\textit{like many other AI-based predictions, our system makes errors}'' after users reported that the system's prediction did not match the intention. However, the author did not explain the potential errors caused by LLMs. \citet{ko2024natural} mentioned that ``\textit{the opaque nature of these models implies that we cannot have full control over their outputs or ensure exact replication in future studies}.''  
Often, authors observed abnormal and inaccurate output from the systems and speculated the reasons for such underperformance. For example, \citet{wang2023enabling} explained the underperformance is that ``\textit{LLMs are trained to generate text instead of the domain-specific task (i.e., selecting an element id in mobile UI)}.'' 
Papers often defer addressing these errors to the future. \citet{buschek2021impact} acknowledged that their work is still a prototype, and suggested quality could be further improved through finetuning or training, ``\textit{possibly involving even larger (email) datasets, extensive architecture search, or generally scaling up.}''

\subsubsection{\limittext{\rr{Resource Limitation}} (28.76\%, N=44)} 
This top-level code refers to computational and financial resources needed to run LLMs, as well as a lack of evaluation standard or metrics. High resource demands can impact the efficiency and scalability of deploying the LLM, and can affect our community's ability to consistently evaluate LLMs or tackle common problems such as hallucination. 

\limit{Computational cost} (9.15\%, N=14):
Computational cost refers to the computational resources required to run LLMs, including the need for hardware (e.g., GPUs) for local execution and limited token windows, which restrict the amount of possible input. For example, \citet{nguyen2024beginning}, who employed OpenAI's Codex, wished that they had used open-source models to ensure study reproducibility, but recognized that doing so would ``\textit{impose significant computational requirements}'' due to the need for extensive GPU resources.
When facing the limited token size, authors had to devise workarounds. 
For example, \citet{petridis2023anglekindling} split their documents into sections to accommodate GPT-3's input length, and wrote that that might have ``\textit{affected the overall performance and user experience of the system}.'' 

\limit{Financial cost} (3.27\%, N=5):
This resource constraint included monetary expenses with using LLMs, often tied to API calls for closed-source models and using online platforms like ChatGPT.
For example, RELIC integrated GPT-3 due to its high performance, but authors also recognized that the LLM-enhanced component via the  API ``\textit{will inevitably increase calculation expenses.}''~\cite{cheng2024relic}
Similarly, financial cost also impacts access to advanced chatbot playgrounds. In a study of ChatGPT's ability to answer programming questions, \citet{kabir2024samia} noted the \$20 per month subscription fee is a ``\textit{considerably high monetary value for many countries},'' and decided to use the free version (GPT-3.5) to lower the barrier for reproducibility at the expense of potential performance.

\limit{Lack of evaluation standards/metrics} (16.99\%, N=26):
This category includes authors wishing to evaluate LLM aspects, but lacking the appropriate standard or metrics. A paper falls under this category only when authors explicitly called for more standards (e.g., ``open question'' or ``active research area''). For instance, \citet{taeb2024axnav} recognized that some participants in their user study spotted errors in their system, but stated that``evaluating the \emph{correctness} of LLM-based systems remain an active area of research.'' 
\citet{cheng2024scientific} mentioned that guardrailing the \emph{safety} of their LLM-powered tool ``without supervision'' in the wild is still an ``active research area''. In the same paper, \citet{cheng2024scientific} recognized that achieving ideal conversational context was still challenging, ``\textit{despite the abundance of literature on effectively engineering prompt for LLMs}.'' 
Several papers also called out a lack of benchmarks for evaluating LLMs outputs, such as conducting thematic analysis~\cite{lam2024concept} and in mental health applications~\cite{kim2024mindfuldiary}.

\subsubsection{\limittext{Research Validity}(90.85\%, N=139)} 
Research validity is often defined as the extent to which an instrument measures what it claims to measure or if the study design can effectively test their hypotheses~\cite{mackenzie2024human}. \textit{Internal validity} refers to the legitimacy of a study's results, considering factors such as group selection, data recording methods, and analysis procedures~\cite{mackenzie2024human}. \textit{External validity} concerns the findings' transferability to other contexts of interest~\cite{mackenzie2024human}. We consider ecological validity a subset of external validity, in that it refers to whether the studies resemble ``real-world'' conditions~\cite{schmuckler2001ecological}. Validity issues can arise across users, contexts, LLMs, and prompts. In total, we identified $2 \times 4$ codes related to this limitation. During coding, we first determined whether the issue impacted internal or external validity, and then identified the affected dimensions. We avoided assessing whether the project could have validity issues, but instead coded what the authors acknowledged in their paper.

The most prevalent limitations are \limit{internal and external validity} \limit{across \emph{users} and \emph{contexts}}. Internal validity issues related to \emph{users} often stemmed from limited sample sizes and lack of diversity within samples. For example, \citet{lin2024rambler} mentioned that ``\textit{a relatively small sample size leads to challenges in concluding some of the potential correlation}.'' This, in turn, may have external validity concerns. For example, \citet{park2024promise} mentioned that their research is based only on English-speaking university students, so the result ``\textit{may not reflect students who speak English as a second language}.'' Similar issues can also apply to different \emph{contexts}, such as application scenarios. \citet{zhang2024mathemyths} recognized that their study setup might have ``\textit{constrained the natural spontaneity that a human can bring to the storytelling process}'', which may have hurt the internal validity of observing behaviors that the authors claimed to study. 
\citet{zhang2024designing} acknowledged that their insights ``\textit{may or may not generalize to use in the field}'', because their prototype design constrained ``\textit{what tasks our participants could do}.'' Research validity issues across users and contexts are generally related to study designs evaluating LLMs or LLM-powered systems.

\emph{Of the 153 papers, 130 papers (84.98\%) used or studied a variation of the closed GPT-family models.}  Despite this, many researchers articulated the \limit{research validity issues across \emph{models}}. Internal validity issues may arise when using LLMs. For example, \citet{chakrabarty2024art} employed the default GPT-4 generation parameters (i.e., temperature = 1) to evaluate the model's capabilities. However, they recognized that a variation in temperature might have changed the content quality, thus affecting the study conclusion. \citet{dang2023choice} also acknowledged that they might not have identified the best settings for model usage due to using the black-box models, which may affect the results internally.

The majority of the research validity issues are around external validity. For example, \citet{dang2023choice} addresses external validity in the same paper, stating that there might be ``\textit{potential changes to the model over time}'', which limit the exact replicability for their studies ``\textit{beyond our control}''. \citet{kobiella2024if} conducted their study with GPT-3.5 and recognized that ``\textit{some findings might not be as prevalent}'' with the release of GPT-4. 

Given this external validity concern, many papers designed their system to be ``\textit{modular}'' on purpose --- swapping the underlying model with other models or even future, non-existing models. For instance, \citet{feng2024mud} mentioned that ``\textit{while we use the gpt-3.5-turbo as our model in the study, we believe that other LLMs trained on similar resources, such as the PaLM and the open-sourced LLama model, could also deliver comparable or even better performance}.'' \citet{goldi2024intelligent} mentioned numerous drawbacks of current GPT-3.5 models but suggested that ``\textit{future improvement in these models could mitigate such limitations}.'' This acknowledgement could potentially defer the responsibility of ensuring research validity in highly context-dependent HCI studies to LLM model developers. 

\limit{Research validity surrounding \emph{prompts}} is another emerging limitation. \textit{Of the 153 papers, 146 conducted some form of systems or studies that prompted LLMs. Of the 146, 40.4\% (N=59) did not release their prompts in the full paper or the supplementary materials.} 
Authors were generally aware that prompt variation could impact their results: \citet{cheng2024scientific} noted that ``\textit{minor prompt adjustments aimed at improving one aspect often had unintended, drastic negative effects on others}.'' 
Similarly, LLM-powered tools, which have been evaluated through technical or user studies, may not generalize externally to other prompts. 
For example, \citet{kabir2024samia} mentioned that the design of the prompt in their study is highly dependent on the questions in their sample. 
Since how people phrase these questions in the real world varies from person to person and situation to situation, more work is needed to evaluate LLMs against prompt variation.
Despite the validity concerns, several authors still proposed to revise the prompt to enhance the system. 
For example, \citet{wang2023enabling} proposed to improve the system quality by adapting their system prompt depending on the input, but acknowledged that this proposal might ``\textit{lead to inferior performances}.''

\subsubsection{\limittext{Consequences} (22.88\%, N=35)} 
This category shows potential negative outcomes 
that may arise from the artifact or study. In some cases, authors present the concerns in an ethics or impact statement (14.38\%, N=22) with concrete remediation strategies. 

\limit{Economic Harms} (11.11\%, N=17)
This refers to potential effects on employment and work. For example, \citet{de2024llmr} highlighted the concern of ``developers and creators being replaced''. However, they also recognized that these tools have not achieved end-to-end development, and if so, the these tools should still require human intervention. \citet{shaikh2024rehearsal} mentioned that their tool to simulate conflict resolution scenarios might cause job replacement and devaluation for expert trainers. Many papers on Communication \& Writing, such as \citet{lee2024design} and \citet{hoque2024the}, stated that their LLM-powered writing tool may change copyright issues and how writers work. 

\limit{Representational Harms} (5.88\%, N=9) This harm refers to social groups being cast in a less favorable light than others, affecting the understandings, beliefs, and attitudes that people hold about these groups~\cite{barocas2017problem}. For example, \citet{benharrak2024writer} recognized that LLM-generated personas ``have the potential to reproduce harmful stereotypes.'' \citet{salminen2024deus} called out that ``as with any novel technology,'' their use of LLM can have adverse societal effects including ``reinforcing gender stereotypes and affecting diversity representation.'' However, these risks were ``not in the scope of'' their study, but warrant further scrutiny from the HCI research community. 

\limit{Misinformation Harms} (2.61\%, N=4) This harm arises from the LLMs outputting false, misleading, non-sensical, or poor quality information~\cite{weidinger2022taxonomy}. For example, \citet{li2024value} added that writers' viewpoints may get misled by ``misinformation generated by AI assistants.'' \citet{tanprasert2024debate} recognized that if they shifted their topic in the study to a more technical topic, which may lead to more cases of LLM hallucination, not only would the research validity have been compromised (i.e., ``the credibility of the information can seriously weaken the chatbot's stance in the study''), but users also may suffer from misinformation spread by the chatbot, if the users are not aware of it. We also found studies that tackle misinformation directly in \citet{zhou2023synthetic} where they examined characteristics of LLM-generated misinformation compared with human creations, and then evaluated the applicability of existing solutions.

\limit{Malicious Use} (1.96\%, N=3) This harm stems from humans intentionally using the LLM to cause harm, e.g., via malware or fraud~\cite{weidinger2022taxonomy}. \citet{precel2024canary} used a whole appendix section to discuss how their findings may harm the Jewish community by anti-Semitic actors. When studying the effect of LLM-powered search systems on information-seeking tasks, \citet{sharma2024generative} recognized that their system and study ``may incur misuse'' because they introduced opinion bias to power the LLM-based search system. Therefore, they ``made public the prompts in the study but will only make them available for requests that we can verify for safe usage (e.g., scientific and non-commercial purposes).'' This approach highlights the interesting balance between ensuring open source and preventing malicious use.

\limit{Hate Speech} (1.96\%, N=3) This category represents prejudice, hostility, or violence against individuals or groups. \citet{de2024llmr} stressed that a serious concern is ``the potential for individuals to generate harmful and inappropriate content'' with their framework, calling for future safeguards. \citet{kim2024diarymate} extensively discussed the ethical concerns of using LLMs for personal journaling. They mentioned that their study may suffer from LLMs' potential ``to generate offensive or violent content." To mitigate this risk, the authors informed participants about potential misbehaviors and offered university mental health care resources in case of adverse events. 

\limit{Environmental harms} (0.65\%, N=1) This category refers to the damage that LLMs can cause the environment, in particular due to the large energy consumption that training and querying requires \cite{weidinger2022taxonomy}. One paper explicitly discussed environmental harms~\cite{lawley2024val}, in the context of worries to scale up their system with LLMs.
\section{Discussion}

\rr{We show substantial growth at CHI in research studying LLMs, echoing trends in other fields~\cite{movva2024topics, wulff_hussain_mata_2024}. 
In this section, we discuss where the CHI community has focused its explorations to-date,  
and what the surge in LLM interest means for HCI's norms around prototyping and design (\ref{sec:dis-opportunities}). 
We then assess issues regarding research rigor that permeate the field (\ref{sec:dis-concerns}). 
We close with a proposal (\ref{sec:dis-questions}): a set of \textit{guiding questions} for HCI researchers to reflect on throughout an LLM-powered project, by considering the \textit{task-appropriateness} of their proposed LLM use, the \textit{validity and reproducibility} of their conclusions, and the \textit{consequences} of their work for research participants, technology users, and society.}

\subsection{Revealed Growth Opportunities for HCI}
\label{sec:dis-opportunities}
To our best knowledge, our work is the first to systematically characterize how LLMs have influenced HCI research.
We find substantial opportunity to expand the application domains where LLMs are used (\ref{sec:dis-hci-applications}), build theories and methods with lasting impact from the large body of empirical work (\ref{sec:dis-hci-contributions}), and standardize how LLMs can and should impact prototyping practices (\ref{sec:dis-hci-prototyping}).

\subsubsection{Beyond language-based applications}
\label{sec:dis-hci-applications}
Our study demonstrates that LLMs are applied across a wide range of \adtext{application domains}, reflecting diversity within the HCI community, and also confirming that LLMs' influence on HCI is pervasive across subareas.
Some areas are well-represented already, and provide examples of how to build new research communities around LLM applications.
Specifically, we found the \ad{Communication \& Writing} domain has garnered the most attention, perhaps due to LLMs' direct relevance to producing language. 
This community has coalesced around such initiatives as the In2Writing workshops at NLP and HCI conferences~\cite{In2Writing}, and \citet{lee2024design}'s effort to chart the design space of intelligent and interactive writing assistants.
Other areas are less represented in our review, and represent opportunities for new research and community-building. For example, papers related to Games and Play were less common in our sample, even though this area is large enough to warrant its own CHI subcommittee.
As LLMs facilitate more games and simulations, we anticipate this area to be a generative site for new work.
Our categorization of application domains can help researchers identify where in the community their interests might fit, and how to develop these areas as LLMs continue to proliferate.

\subsubsection{Beyond empirical and artifact contributions}
\label{sec:dis-hci-contributions}
While we observed that HCI researchers succeeded in applying LLMs to a diversity of application domains, we found less diversity in the \cttext{contribution types} pursued in the literature.
The LLM-related papers in our sample predominantly center on artifact contributions and empirical evaluations, often in the form of user studies of new artifacts. 
Empiricism is central to understanding phenomena; however, to develop knowledge from our aggregate body of observations, we encourage more attention to \citet{wobbrock2016research}'s five other contribution types, each of which was less well-represented in the literature.

We observed an opportunity for the community to further pursue \ct{dataset contributions}~\cite{wobbrock2016research}---and approaches to data collection that center real user needs and downstream harms.
Traditional NLP benchmarks are often criticized for their lack of \textit{context realism}: the model performance measures are often divorced from downstream use cases \cite{liao2023rethinking}. 
Adopting community-driven and participatory approaches to benchmarking could provide data that represents real and diverse user requirements, while still enabling developers to test LLMs' capabilities \cite{suresh2024participation}. HCI's sociotechnical approach is well-positioned to innovate on benchmarking culture: e.g., Bragg et al. have explored how to build automatic sign language translation tools by crowdsourcing video datasets with the Deaf community \cite{bragg2022exploring, tran2023us}, Reinecke et al. have developed volunteer-based online platform to reach larger and more diverse user population~\cite{reinecke2015labinthewild,reinecke2013perdicting}, and
\citet{kuo2024wikibench}'s Wikibench offers a community-driven alternative to data curation that captures community consensus.

We also see significant need for theoretical and methodological contributions, as well as \rr{literature review} and opinion pieces, all of which can help shape public perception and understanding of LLMs' pitfalls and potential.
\ct{Theoretical} and \ct{methodological contributions} can draw transferable principles from bodies of empirical and artifact research~\cite{bellotti2002making, wobbrock2016research}. Based on our review, the field would benefit from more work on, e.g., the design space around LLMs in the various application domains (cf. \citet{lee2024design}'s work on intelligent writing assistants), or the design processes behind developing LLM-based systems. 
\ct{Literature review} like the present study and \ct{opinion contributions} can also help us reflect on our community's progress and help the field to identify and address emergent issues (cf. \citet{correll2019michael}'s work advocating for moral obligation among researchers and designers in visualization).
Our literature review can help fill this void, but more work is needed. 
Targeted literature review on more specific topics can further guide our community forward, and we especially encourage papers that synthesize lessons bridging HCI and fields like NLP and ML.

\subsubsection{How LLMs impact prototyping standards}
\label{sec:dis-hci-prototyping}

More broadly, our findings signal broader methodological questions for HCI:
What level of prototype fidelity is needed to demonstrate a new interaction---and relatedly, what level of system-building and evaluation is needed to make an artifact contribution? This question arises from our challenge to define which papers proposed \ct{artifact contributions}, and which used \role{LLMs as system engines}.
Throughout HCI, \textit{Wizard of Oz} approaches have long been used to prototype interactions with intelligent agents \cite{maulsby1993prototyping, dahlback1993wizard}. 
These methods typically present a research participant with an interface that appears to have machine intelligence, but unbeknownst to them, a human performs those functions (cf. \cite{gu2024analysts}).
Wizard of Oz approaches gained popularity in HCI as methods that allowed rapid and inexpensive prototyping of future technological capabilities.

However, the utility of these methods may change as LLMs proliferate.
If a researcher explores the design space around using LLMs in a given domain---to ``\textit{sketch with AI}'', as \citet{yang2019sketching} describe---does a Wizard of Oz approach provide benefits over a fully automated approach anymore?
Historically, researchers have been trained to prototype quickly and cheaply, and thus they might conclude that a Wizard of Oz approach makes more sense.
Today, however, LLMs have likely lowered the barrier of developing systems so much that we may expect designers to use them to achieve more ecologically valid research. After all, even the best of Wizard of Oz methods cannot perfectly proxy machine intelligence \cite{yang2020re}.
If a designer creates an LLM-backed prototype, however, what level of performance should they aspire to in their system? Should the prototype be evaluated as an artifact contribution? Is it a system contribution, if the implementation is straightforward, given LLMs' capabilities? 
For HCI, this debate has ramifications not only for the methodological norms we use, teach, and expect in peer review, but also for the research validity that we produce and the contribution that we value.
We encourage the HCI community to collectively reflect on what widespread LLM usage means for prototyping standards, and the resulting implications for how HCI produces knowledge.

\subsection{Challenges: Validity, Reproducibility, and Consequences}
\label{sec:dis-concerns}

To achieve the opportunities we outline in the previous section, the HCI community will also need to reflect on some fundamental challenges with LLM research identified in our analysis.

\subsubsection{Proprietary LLMs raise reproducibility concerns}

Our analysis showed that despite authors' commonly articulated limitations surrounding \limittext{research validity}, 
papers using LLMs is growing exponentially. This trend adds urgency to calls for examining reproducibility in HCI~\cite{cockburn2018hark, salehzadeh2023changes}.

We found that an overwhelming 
majority of our sample (84.98\%) studied a variation of the closed GPT-family models ($N_{\texttt{GPT-4}}=61, N_{\texttt{GPT-3.5}}=41, N_{\texttt{GPT-3}}=26$).  
Using closed models can pose serious problems for research reproducibility, especially if authors choose not to disclose prompts.
Researchers have shown that proprietary and closed models, often accessed through APIs, are constantly changing, and may even inject unspecified edits to a user's prompt (e.g., system prompts)~\cite{rogers-etal-2023-closed}, meaning model behavior may be unpredictable despite using the same, disclosed prompt.
Proprietary models also generally do not disclose their training data or model weights, which makes understanding model behaviors and properties in applications and downstream systems difficult. Most papers also did not justify their model choice, though model families can exhibit quite different behaviors. For example, models optimized for chat and those optimized to take instructions can be expected to behave differently~\cite{zhang2023instruction, ouyang2022training}. In our study, we found few authors explicitly justified their model choice for their use case. 
Some artifact contribution papers implied that their systems could be \textit{modular}, suggesting the LLMs in the system could be customized by users or in future work. If an LLM-powered system is meant to be modular, and models exhibit different behaviors, then developers should disclose and discuss how model choices might affect the system for future users and developers.
To further complicate the issue, 40.41\% of the papers ($N=59$) did not release the prompts in the paper or supplementary material. Given LLMs are sensitive to subtle changes in prompt formatting 
even in large models~\cite{sclar2023quantifying}, the lack of transparency in prompt design and usage may affect system performance, and prevent researchers from replicating and building upon existing work.

\subsubsection{LLM properties introduce additional research validity concerns}
Our analysis surfaced the fact authors have many concerns around how LLMs' inherent properties might impact research validity---but less knowledge about what precisely to do about it. 
Whether LLMs were the object of study or powered a system with which users interact, researchers readily acknowledged issues like \limit{LLMs' inherently limited training data}, penchant for \limit{hallucination}, and \limit{nondeterministic responses}.
Some of these limitations shaped whether we \textit{should} expect certain behaviors from LLMs (e.g., limitations on training data), and others shaped whether research results could be considered externally valid (e.g., hallucination and nondeterminism).

However, we found that the most commonly mentioned LLM-related performance limitations were \limit{unspecified errors} \limit{and biases} (16\%, N=24). Though authors have some awareness of LLMs' limitations, this code's prevalence indicates that further engagement with the precise nature and performance effects of these errors was often unaddressed. 
Being more specific about the nature of errors or bias arising from LLMs and how this may affect the system or results is critical for a reader's understanding of the nature and extent of the stated limitation. For example, the more specific issues captured by other codes, e.g., ``hallucination,'' bring with them the ability to better interpret specific potential failure types of a system, and even imagine potential downstream harms. 
We urge HCI researchers to more precisely specify what potential errors and biases they identify in their use of LLMs, so that consumers of our research can better understand how the systems built upon these technologies may fail.

\subsubsection{Consequences, Risks, and Broader Impacts} In parallel to the \textit{limitations} around validity and reproducibility described in the previous section, we found tremendous need to confront how HCI researchers assess and report the \textit{consequences and risks} of their work.

First, we explicitly differentiate between \textit{limitations} and \textit{consequences}. \textit{Limitations} refer to factors that affect the truthfulness of the paper's conclusions, such as issues with validity, transferability, and generalizability. \textit{Consequences} pertain to long-term social impact, including insights that could help guide real-world deployments.
In fields such as ML/AI and computer security, recent initiatives have asked authors to provide \textit{ethics statements}~\cite{hecht2021itstimesomethingmitigating}, \textit{broader impact statements} \cite{nanayakkara2021unpacking}, and other structured ways of reflecting on the consequences of their work.

While authors considered questions of validity and reproducibility---limitations of their work---only 35 papers discussed potential consequences of their findings and results, often in an ethics statement (N=22).
Ethics statements were discussed among HCI researchers in 2018~\cite{hecht2021itstimesomethingmitigating}, but to-date have not been formally standardized in CHI's submission process; however, they have been used in ML and AI conferences including NeurIPS and FAccT \cite{nanayakkara2021unpacking}.
As our study showed that LLMs have been used in diverse applications and are changing research practices, we believe that the CHI community should place greater emphasis on discussions around consequences. Encouraging a more explicit discussion ensures that the HCI contributions are responsibly aligned with the broader societal good.

More broadly, we contend that structured consideration of consequences --- via an ethics statement or other means --- would help HCI lead the scientific community by demonstration as LLM-based work proliferates. 
As HCI research inherently considers people and society, its innovations are likely to be deployed and have impact with real-world users~\cite{lucas2018translational}.
Establishing field norms to consider consequences can help HCI lead in engaging with LLMs in rigorous and thoughtful ways, providing a model for researchers and practitioners across disciplines.
Below, we contribute an initial proposal we hope accelerates the scientific community towards this vision.

\subsection{Guiding questions for HCI researchers using LLMs}
\label{sec:dis-questions}

Given serious concerns around validity, reproducibility, risks and consequences (\ref{sec:dis-concerns}), how can we move towards the opportunities outlined in \ref{sec:dis-opportunities}?
As a first step, we contribute practical guidance to prompt researchers' reflection on the \textit{validity} and \textit{appropriateness} of their LLM-related work.
We view LLM-related research not as categorically harmful, but rather that using LLMs requires careful consideration throughout a project.

\rr{In this section, we draw on the \roletext{LLM roles} and \limittext{Limitations} to synthesize \textit{guiding questions} for researchers and practitioners to consider at each stage of the research process, and for each role that an LLM might take. 
Our questions are intended as prompts for reflection, not a checklist for completion, and should be used iteratively to ensure that any work continually centers thoughtful LLM usage.
We chose more open-ended questions over prescriptive guidance to uplift critical thinking around LLM usage in research design.}
\generalLabel{G} represents \textit{general} questions for any project, whereas \sublabel{S} refers to \textit{specific} questions for each role.

\textbf{1. \generalLabel{G} What role will the LLM play in your project?} \rr{A researcher should understand the stage at which an LLM might be included (\autoref{fig:llm-role}).} A key question is whether an LLM is needed at all---whether achieving the same result is possible using alternate approaches that are better established or less costly, such as using simpler models or humans. 

\textbf{2. \generalLabel{G} Which model is appropriate?} This question helps authors decide between open and closed models. 
As \citet{palmer2024using} discuss, closed models are at odds with the transparency and reproducibility expected of research. 
Using closed and proprietary LLMs may also violate study ethics if participants have not consented to sharing data with the LLM. 
Still, closed models may be appropriate if, e.g., LLMs are the object of study (as in an audit study); if cheap and rapid prototypes are needed; or if a closed model was shown to be the state of the art in a specific task, and is used only for that task. If others may treat the LLM in a system as modular, then consider the robustness of the chosen model and the impact of swapping in  different models. 
Researchers should consider such factors and justify their model choice. 

\textbf{3. \generalLabel{G} How did you disclose the models and prompts?} This question encourages authors to document model versions and prompts used in their study. For models, we encourage specificity: e.g., \texttt{gpt-4o-2024-08-06} and \texttt{gpt-4o} can refer to different models. Authors should also clearly document the full prompts, or the prompt templates provided to users. \rr{Authors can also consider other methods to improve the research validity, e.g., fine-tuning an open source models on domain-specific dataset~\cite{wei2021finetuned, hu2021lora}.}

\textbf{4. \generalLabel{G} What are the potential limitations of using LLMs for your selected role?} For each LLM role, we contribute specific sets of reflective questions. 

\begin{itemize}
    
    \item \role{LLMs as system engines}: 
    \begin{enumerate}[label=\sublabel{S}]
        \item \textit{What level of artifact fidelity is appropriate to support the contribution?} If the main contribution is a formative study or user perceptions of specific LLM outputs, enabling the interaction is perhaps more important than deploying a fully-functional system. A Wizard of Oz approach may be more appropriate than building a system around a commercial LLM API.
        \item \textit{How would factors like models and prompts affect the system performance?} This clarifies whether the system can achieve the claimed effectiveness with different models or changes in prompts.
        \item \textit{How would factors like models and prompts used in the system affect the research validity of the user study?} Authors should consider whether the LLM-powered system is robust across users, and note any discrepancies between target system users and recruitment population.
    \end{enumerate}

    \item \role{LLMs as research tools}: 
    \begin{enumerate}[label=\sublabel{S}]
        \item \textit{Why are LLMs appropriate for your research task?} 
        If your task is \textit{classification}, e.g., labeling a dataset, using an LLM may overlook nuances in the human \textit{perspectives}.
        If your task is \textit{generation}, e.g., creating survey questions or datasets, using an LLM risks neglecting lay and domain \textit{expertise}.
        \item \textit{How can you evaluate the performance of your LLM-based research tool?} 
        Across tasks, validation via human or formal methods is often needed to quality-check an LLM's outputs. 
        These evaluations are vital, but the human effort needed to structure and faithfully execute them may exceed the utility of using the LLM in the first place---what \citet{bainbridge1983ironies} calls the ``automation trap''.
        \item \textit{How will the performance of your LLM-powered research tool affect the validity of your research?} In addition to ensuring research tools remains standardized and accurate, authors should understand how the choice to use an LLM would affect the claims of the empirical work.
    \end{enumerate}

    \item \role{LLMs as participants \& users}. Consider the questions under \role{LLMs as research tools} above. Then, specifically: 
    \begin{enumerate}[label=\sublabel{S}]
        \item \textit{Given LLMs' known inability to faithfully represent people, how can an LLM-powered tool adequately stand in for the target population in your study?} 
        Using LLMs to simulate users 
        deprives them of the opportunity to consent to such research \cite{agnew2024illusion}. 
        LLMs also run the risk of misrepresenting people and are unlikely to faithfully portray identity groups due to the nature of their training data \cite{wang2024large}.
        Given these known constraints, consider how to adjust your study design to enable people from your target population to evaluate the LLM's outputs, and determine how they are used (cf. \cite{tan2024more, suresh2024participation}). 
        Throughout, stay attuned to whether the effort required for proper human evaluation and participation exceeds the benefits of introducing an LLM in the first place.
        \item \textit{Given that they are only trained on human language, how can LLM-backed tools reflect the realism of human behavior and opinion dynamics of interest?} Although LLMs might display human-like behaviors and opinion dynamics by modeling language, they often struggle to generate outputs that capture the complexity and diversity of real human interactions shaped by individuals' lived experiences. Human language is also inherently limited in capturing the fidelity of human behavior, which can threaten this method's validity.
        Given these known constraints, consider how to adjust your study design to evaluate the LLM's outputs against real discourse and deliberation (cf. \cite{kuo2024wikibench}). 
        Again, consider whether the effort required for proper human evaluation might exceed the benefits of introducing an LLM.
    \end{enumerate}
    
    \item \role{LLMs as objects of study}: 
    \begin{enumerate}[label=\sublabel{S}]
        \item \textit{What model behaviors do you aim to study?} When studying LLMs directly, authors should consider a common feature of most LLMs, a feature of particular model families, or just one particular model (e.g., \texttt{gpt-4o-2024-08-06}).
        \item \textit{How did you ensure that the claims made about the models were appropriate?} Authors should consider whether the findings can generalize to other models. If not, authors should not overclaim the findings.
    \end{enumerate}

    \item \role{User perceptions of LLMs}: 
    \begin{enumerate}[label=\sublabel{S}]
        \item  \textit{Who are the representative participants for the study?} Authors should consider how their participants impact the \textit{internal validity} of their work: whether their study sample accurately reflects the population they claim to represent.
        \item \textit{What confounds could impact participants' perceptions of LLMs?} LLMs are subject to tremendous hype in the popular press. Participants may come with preconceptions about LLMs' capabilities that require researchers' attention. For example, a participant who has seen ads from AI companies may more quickly grasp the affordances of a new LLM-powered interaction paradigm than a participant who does not experience AI filter bubbles.
    \end{enumerate}
    
\end{itemize}

\textbf{\generalLabel{G} 5. What are the potential \limit{consequences} of your study?}
LLMs have known \textit{environmental} costs authors should consider in the study design (cf. \cite{luccioni2024power}).
Having participants interact with LLMs may also impact \textit{privacy} \cite{carlini2022quantifying}, especially when using closed models; thus authors may consider how to obtain consent for an LLM to use a participant's data, how to sanitize LLM inputs, and measures to protect participants' agency over their data.
HCI researchers studying LLMs---especially when they augment or replace human effort---should consider the systems' \textit{economic} impacts.
LLMs' need for massive datasets can create global inequalities for data workers \cite{gray2019ghost}; and companies may prioritize investing in LLMs over workers, even as humans are needed to ensure LLMs function properly \cite{bainbridge1983ironies}. 

\subsection{Limitations}

Our work has several limitations. First, our sampling approach might not cover all papers that used LLMs. For instance, we found one paper in our robustness check that mentioned GPT-4 just once, in their methods, without mentioning any other keywords in our list. \rr{Other works may have even used LLMs in their methods without mentioning them at all, which would align with the increasing interest in using LLMs to automate academic research~\cite{lu2024aiscientistfullyautomated}. Our work primarily focused on prompting as the main interface, but future study may extend our samples to study and identify best practice for other techniques (e.g., fine-tuning~\cite{han2024parameter}, LLM-based embeddings~\cite{petukhova2024text}, and multi-agents~\cite{guo2024large}). While insights from this paper (e.g., computational cost) remain relevant, additional research validity concerns may emerge, e.g., challenges in sharing datasets to replicate fine-tuning results or agent configurations to reproduce multi-agent system outcomes.} 

Second, our review was limited in scope by the manual and iterative process. 
Using LLMs to conduct analyses like ours is an active research topic and can increase scale, but we chose not to use LLMs because of concerns with using LLMs without proper disclosure and evaluation. In our preliminary phase, we used \texttt{gpt-4o-2024-05-13} to explore paper topics, but found the themes too general to gain meaningful insights. To use LLMs effectively, human qualitative coding will likely still be required to develop effective prompts and validate the accuracy of the LLM classifier. 
Hence, we spent hours curating the dataset, reading papers, resolving coding disagreements, and discussing difficult papers. The laborious nature of this process prevented us from conducting a more expansive literature review.
Future researchers might use our paper as a starting point to examine LLMs' impact on other HCI subcommunities and even conferences in other fields, e.g., by reviewing the (dis-)connection between the NLP and HCI communities. 

\bibliographystyle{ACM-Reference-Format}
\bibliography{sample-base}


\end{document}